\documentclass[prb,twocolumn,showpacs]{revtex4}

\usepackage[german,english]{babel}
\usepackage{amssymb}
\usepackage{amsfonts}
\usepackage{amsmath}
\usepackage{mathrsfs}
\usepackage{graphicx}

\begin{document}

\title{Sensing DNA -- DNA as nanosensor: a perspective towards
nanobiotechnology}

\author{Ralf Metzler}
\email{metz@nordita.dk}
\affiliation{NORDITA - Nordic Institute for Theoretical Physics,
Blegdamsvej 17, DK-2100 Copenhagen {\O}, Denmark}
\author{Tobias Ambj{\"o}rnsson}
\email{ambjorn@nordita.dk}
\affiliation{NORDITA - Nordic Institute for Theoretical Physics,
Blegdamsvej 17, DK-2100 Copenhagen {\O}, Denmark}

\date{\today}

\begin{abstract}
Based on modern single molecule techniques, we devise a number of possible
experimental setups to probe local properties of DNA such as the presence
of DNA-knots, loops or folds, or to obtain information on the
DNA-sequence. Similarly, DNA
may be used as a local sensor. Employing single molecule fluorescence methods,
we propose to make use of the physics of DNA denaturation nanoregions to find
out about the solvent conditions such as ionic strength, presence of binding
proteins, etc. By measuring dynamical quantities in particular, rather
sensitive nanoprobes may be constructed with contemporary instruments.
\end{abstract}


\maketitle

\section{Introduction}

Single molecule techniques allowing both the manipulation and probing of
single molecules, have come of age. Optical tweezers, atomic force
microscopes, or single molecule tracking and optical detection methods (for
instance, fluorescence correlation spectroscopy, FCS, or fluorescence
(F{\"o}rster) resonance energy
transfer, FRET) have become standard methods in laboratories. By means of
these techniques having access to scales in the nanometre domain allows us
to obtain quantitative
information about the physical properties of molecules without being
masked by the inevitable ensemble averaging inherent in bulk
measurements. Even though typical single molecule data are more noisy than
bulk signals, the gain of individual molecular behaviour by far
outweighs this disadvantage. In certain cases, single molecule experiments
can reveal information, that is not accessible to bulk measurements, for
instance, the recent experiments on the characteristics of single-stranded
DNA-binding proteins \cite{pant}, or the measurements of the passage of
single biopolymers through nanopores \cite{john,amit}. Moreover,
one may even extract information from the single molecule noise; for example,
on the nature of such known phenomena as Brownian motion \cite{kirstine}.
This progress is essential to recent advances in a
number of fields like biological and soft matter physics, or
nanobiotechnology.  The small system sizes also make it possible to test
fundamental physical theories such as the Jarzinsky relation
connecting measurements of the nonequilibrium work needed, e.g., to
stretch an RNA segment \cite{liphardt}, to the difference in the
corresponding thermodynamic potential
\cite{jarzynski}; or the entropy production along
single trajectories exposed to stochastic forces \cite{seifert}.

In what follows, we devise a number of potential experimental setups
probing on scales down to the nanolevel, both the physical behaviour
of DNA itself as well as different ways to employ DNA as a
nanosensor. A certain emphasis is put on methods where theoretical models
are available so the physical parameters of the DNA and its surroundings may
be {\em quantitatively} extracted from experimental data. These setups
should be well within reach of the state of the art techniques and
may be used to obtain important new information on DNA, or prompt
new technologies based on DNA. As the DNA molecule is the main
ingredient for our exposition, we start with a primer on the physical
properties of DNA, before embarking for setups to probe (some of)
these properties on the single molecule level and propose several
possibilities to use DNA as a sensor.

\section{DNA-physics}

DNA has a number of remarkable properties. Made up of two chemically very stable
individual molecules that wind around each other to produce the
double-helix, it carries, embedded in its core, the entire genetic code
of an organism. Modern gene technology is able to produce custom-designed
DNA molecules with any given sequence. There exists proteins ("biological
glue") by which DNA can be attached to microbeads, that, in turn, can be
manipulated by optical tweezers or microbeads. These properties make DNA
an ideal object for single molecule experiments.

\begin{figure}
\includegraphics[height=7.6cm]{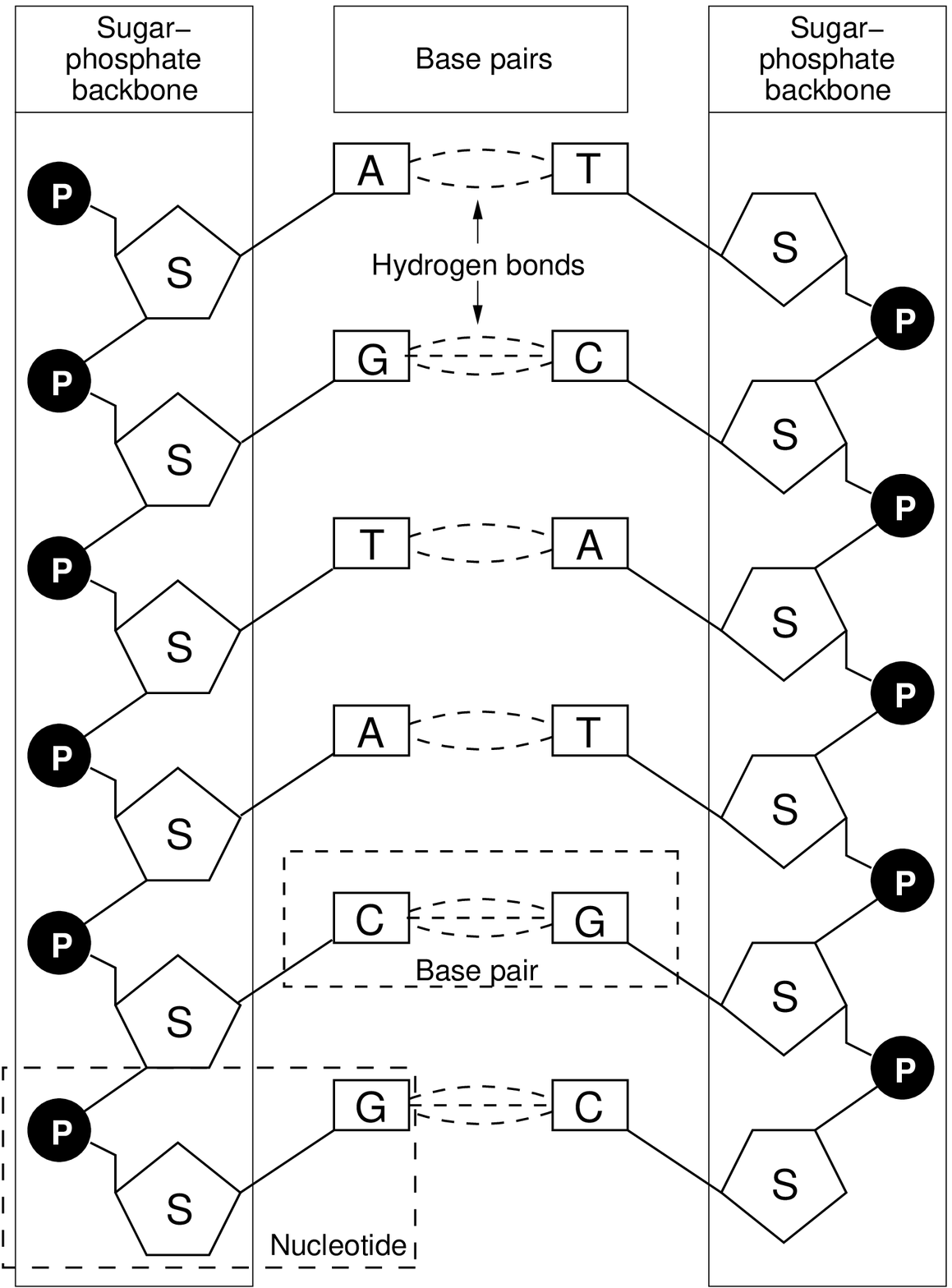}
\hspace*{0.4cm}
\includegraphics[height=7.6cm]{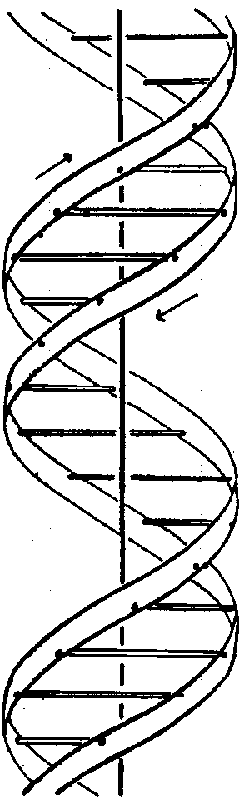}
\caption{Left: Schematic view of the chemical structure of the DNA molecule,
showing the bases suspended by the outer {\bf S}ugar-{\bf P}hosphate scaffold.
Right: Reproduction of the original graph of the proposed double-helical
structure of DNA. Reprinted with permission from Reference
\protect\cite{crick}, Watson and Crick, {\emph Nature} 171, 737 (1953).
$\copyright$ 1953, \emph{Nature}.}
\label{dna}
\end{figure}

DNA consists of a backbone of sugar and phosphate molecules suspending
the base-pairs in its core, see Figure \ref{dna}. This ladder structure
in 3D forms the spiral staircase structure (see Figure \ref{dna} on the
right) originally predicted by Watson and Crick
\cite{crick}. The Watson-Crick double-helix, or, more
precisely, its B-form, is the thermodynamically stable configuration
of a DNA molecule under physiological and a large range of in vitro
conditions. This stability is effected first by Watson-Crick H-bonding,
that is essential for the specificity of base-pairing ("key-lock principle").
Base-pairing therefore guarantees the high level of fidelity
during replication and transcription. The second, major, contribution to
DNA-helix stability comes from base-stacking between neighbouring
base-pairs, through hydrophobic interactions between the planar
aromatic bases, that overlap geometrically and electronically
\cite{delcourt,kornberg}.

The relevant length scales of DNA span several orders of magnitude
\cite{kornberg,frank,frank1,calladine}. The distance between
neighbouring base-pairs is approximately 3.4 {\AA}, while the hard
core diameter of DNA is 2 nm. One full turn of the double-helix is
made up of 10.5 base-pairs. The persistence length, i.e., the distance
over which the tangent-tangent correlations decay, is of the order of
50 nm (340 base-pairs), more than an order of magnitude larger than the
diameter. Locally, double-stranded DNA (dsDNA) therefore appears
stiff. In contrast, single-stranded DNA (ssDNA) has a persistence
length of a few nm, depending on solvent conditions and sequence. Finally, the
overall length of naturally occurring DNA ranges from several $\mu$m in
viruses, over some mm in bacteria, to tens of centimetres in higher
organisms. The South American lungfish hosts 35 m of DNA per cell
\cite{kornberg}.

An important feature of double-stranded DNA (dsDNA) is the ease with
which its component chains can come apart and rejoin, without damaging
the chemical structure of the two daughter-strands. This unzipping of
the H-bonds between base-pairs is crucial to many physiological
processes such as replication and transcription.
Classically, the melting and reannealing
behaviour of DNA has been studied in solution in vitro by increasing
the temperature, or by titration with acid or alkali. Such
equilibrium measurements are described by the Zimm-Poland-Scheraga
model based on the following physical parameters of DNA
\cite{zimm,Poland_Scheraga,wartell,richard}:
(i) the statistical weight $u=\exp(-\beta\epsilon)$ (with $\beta=1/(k_BT)$,
where $k_B$ is the Boltzmann constant, and $T$ the temperature),
associated with the free energy $\epsilon$ of breaking a single base-pair.
Note that $\epsilon$ is smaller for AT than for GC bonds
\cite{kornberg,delcourt,blake}. $u$ also depends on ambient
salt concentration, applied torques and forces; (ii) the
non-universal prefactor $\sigma_0\ll 1$ that measures the loop initiation
energy associated with breaking the stacking interactions
\protect\cite{wartell,blake,blossey,Poland_Scheraga}; (iii) and the
loop closure exponent $c$ that stems from the entropy loss due to the
closed loop structure of the ssDNA bubble,
compare \cite{Poland_Scheraga,blake,richard}.

While the double-helix is the thermodynamically stable configuration
of the DNA molecule below the melting temperature (or at
non-denaturing pH), even at physiological conditions there exist local
denaturation zones, so-called DNA-bubbles, predominantly in AT-rich
regions of the genome \cite{wartell,Poland_Scheraga}.
A DNA-bubble is a dynamical unit, whose size varies by
thermally activated zipping and unzipping of successive base-pairs at
the two zipper forks where the ssDNA-bubble meets the intact
double-helix. This DNA-breathing is possible due to the fact that on
bubble formation the enthalpy cost and entropy gain, despite each
being significant amounts in terms of $k_BT$, almost cancel and the
unzipping of a base-pair involves a free energy cost of the order of a
$k_BT$. 

We will in the subsequent sections discuss different possible
experimental setups that allow for the measurement of the properties
of DNA and its surroundings.

\section{Sensing DNA: Nano-setups measuring the physical properties of the
molecule of life and its environment}

In this section, we propose a number of arrangements by which physiological
processes and the fundamental physical properties of DNA can be monitored.
Apart from measuring the characteristics of DNA itself, micro- and nanosetups 
are suggested for obtaining information about its topological state or the
solution conditions.

\subsection{Melting and monitoring a nanoregion of DNA}
\label{memo}

\begin{figure}
\includegraphics[width=8cm]{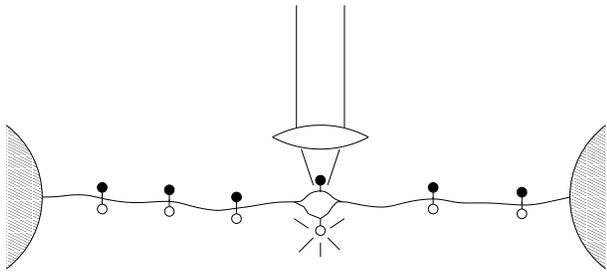}
\caption{A local denaturation zone can be detected by a microscope through
the fluorescence of a dye. Conversely, the DNA being closed at the position
of a fluorophore-quencher pair, the close proximity of the dye to the (black)
quencher prevents fluorescence. Various ways of externally inducing the bubble
are discussed in the text.}
\label{localmelt}
\end{figure}

The local stability of DNA can be probed as sketched
in Figure \ref{localmelt}.
Here, a linear stretch of DNA is held in place by two microbeads,
and a local denaturation zone is monitored by fluorescence of a fluorophore at
the bubble position, for instance, by fluorescence correlation spectroscopy
\cite{altan}. Recent
developments in the theoretical description of DNA breathing
dynamics \cite{PRL,JPC,abm,hetero} relate measurable
dynamical quantities to the Zimm-Poland-Scheraga physical parameters
discussed in the previous section, as well as to the properties of the
surroundings \cite{PRL,JPC}. In particular the fluorescence correlation
could be quantified and shown to depend on (i) the local statistical
weights $u$, i.e, temperature, salt concentration, twist, as well as
the local DNA sequence; (ii) the bubble initiation parameter
$\sigma_0$; (iii) the loop exponent $c$; (iv) the concentration and
binding constants of single-stranded binding proteins; (iv) the rate
constant for unzipping and unbinding, respectively. In addition, the
presence of double-stranded binding proteins could be detected through
the relaxation time spectrum \cite{hetero}.

Alternatively to probing spontaneous DNA-breathing due to thermal
fluctuations, a bubble can also be induced by mechanical stretching of
the DNA, and then the fluorescence traces along the DNA could successively
reveal regions of high and low AT-content. It could also be measured
how occasional
multiple bubble states develop, for instance, how bubbles coalesce
across a GC-rich barrier between two AT-rich bubble domains. Finally,
bubbles might be induced at a selected location observed by microscope
through a strong laser beam or with confocal light of a different wavelength.
This technique may in fact be employed to DNA-sequencing, distinguishing
AT-rich regions from GC-rich, analogously to bulk melting experiments on
the basis of which coding regions of the DNA could be identified
\cite{yeramian}.

We also mention a potential measurement of DNA-mechanics
based on the different persistence length between dsDNA and ssDNA
connected to this setup. Namely, by inducing a larger region of DNA to
denature, one reduces the local stiffness, and by this also the average
resistance of the DNA to longitudinal tension. A change in temperature or
the presence of a denaturing agent should, in principle, be visible through
an increase of the extension between the two microbeads.

\begin{figure}
\includegraphics[width=8cm]{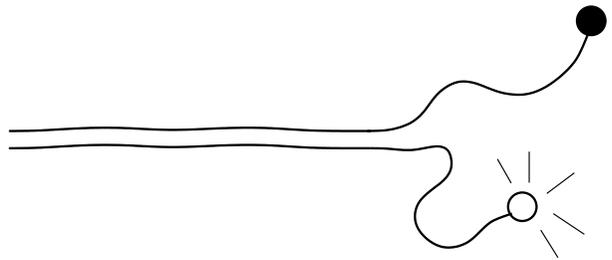}
\caption{Denaturation beacon setup. The right part of the DNA is designed
to be rich in AT,
and preferentially opens up from its ends. The left part, rich in GC or
equipped with an end-loop, stays closed. Once open, the fluorophore-quencher
pair is separated, and fluorescence starts.}
\label{beacon}
\end{figure}

\subsection{Denaturation beacon as sensor}

Unclamped DNA preferentially opens up at the ends (see Figure \ref{beacon}),
as this does not
involve the typical energy barrier for bubble initiation in the middle
of the DNA \cite{Poland_Scheraga}. Having a DNA construct that is rich in AT at
one end and rich in GC at the other end (or that has a closed loop at
that end) could then serve as a molecular beacon sensing the solvent
conditions in small volumes, for example, in gene microarrays. The dynamics
of an ionic
fluorophore-quencher pair depends on the statistical weight $u$, and
one would thereby have a rather sensitive probe for measuring (i) the
presence and concentrations of (multivalent) ions in solution; (ii)
the presence of single-stranded binding proteins; or (iii) local
temperature gradients.

\subsection{Monitoring replication and transcription progress}

\begin{figure}
\includegraphics[width=8cm]{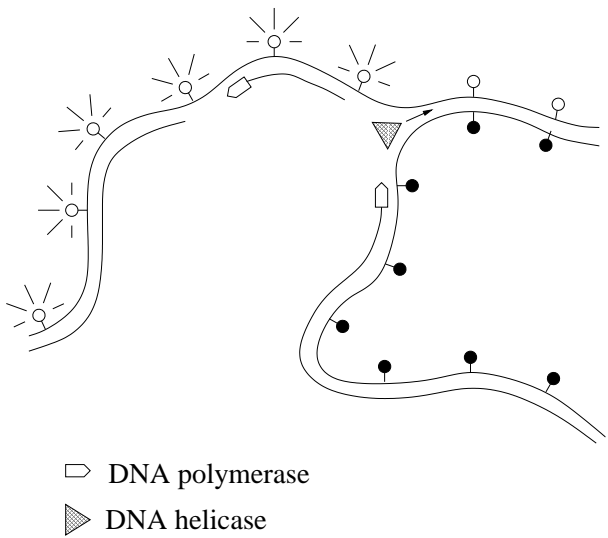}
\caption{DNA during replication by DNA helicase and DNA polymerases. The
DNA molecule is lined with fluorophore-quencher pairs that start to
fluoresce once they are separated spatially.}
\label{poly_monitor}
\end{figure}

Figure \ref{poly_monitor} displays
a DNA-molecule that is being replicated by action of DNA helicase and
polymerase \cite{kornberg,kornberg1}. The DNA molecule is lined with
fluorophore-quencher pairs. Close together in intact double-strand,
fluorescence is quenched; once separated during the replication or
transcription processes, fluorescence occurs \cite{olegrev}.  The
position of the replication fork along the DNA can be monitored
similar to a radar trace either by microscope or digital camera. Using
dyes that bleach out on an appropriate timescale, the observed
fluorescence occurs only in close vicinity of the helicase
molecule. To enable a reference frame for the motion, the molecule
can be held in place by optical tweezers, as used in some of the other setups
in this study.

The setup in Figure \ref{poly_monitor} may be used to measure the {\em local}
transcription/replication speed. The local speed will depend on (i)
the energy needed to break a bond, i.e. the local statistical weight
$u$ (which in turn depends on, for instance, salt concentration); (ii)
the presence of a knot or a kink in the DNA, which would decrease the
local speed (compare \cite{deibler}); (iii) the presence of double-stranded
binding proteins
would slow down or completely halt the opening at the replication
fork; (iv) single-stranded binding proteins would possibly help in the
unzipping process and thereby increase the local speed.
Furthermore, in combination with twisting by magnetic tweezers, over-
or underwound states can be created, and the interplay of
transcription or replication speed with twist or twist-induced
superstructure studied.

\subsection{Locating a DNA knot and measuring its size}\label{sec:DNA_knot}

DNA-knots are created physiologically, and can be detected and removed
by certain enzymes \cite{nanohand}. A number of questions about such
knotted states of DNA are still unresolved, for instance, how a knot can
be detected by topoisomerases; how a knot reduces the transcription speed;
how much it decreases the rupture strength of DNA; or, whether knots at
sequence-determined or chemically stabilized positions are relevant in
gene regulation by bringing segments of the DNA that are distant in the
chemical coordinate along the DNA backbone, close to each other in physical
space.

\begin{figure}
\includegraphics[width=8cm]{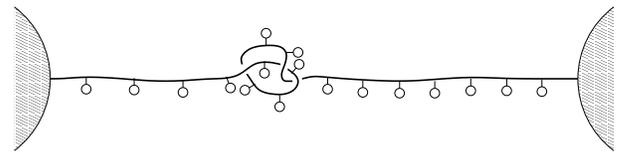}
\caption{Fluorescent labels along a knotted DNA will show increased local
intensity at the knot position. By monitoring the size of the brighter
spot, and its position, important information can be obtained about the
effects caused by the knot state. In particular, sequence dependence and
the influence of stabilising ligands can be studied.}
\label{localknot}
\end{figure}

The setup shown in Figure
\ref{localknot} allows for the measurement of the local brightness of
fluorescence labels along the DNA, which in turn allows for the
determination of: (i) the position of the knot; experimentally, a knot
in a DNA lined with fluorophores can be monitored through increased
local fluorescence where the knot entangles a portion of the
DNA-molecule; a first knot observation study using homogeneous staining of the
DNA was reported recently \cite{quake}. (ii) the local properties of the knot;
by releasing or
increasing tension along the DNA, it can, for instance, be monitored,
whether the size of the knot changes, or whether it is always
tight. Again, effects of additional twist, sequence, and solution
conditions can be probed.

One might replace the fluorescent labels by plasmon resonant
nanoparticles \cite{Schultz} or quantum dots \cite{dot}.
Such particles have the advantage of not photobleaching. Furthermore,
plasmon resonant particles, that are in close proximity to each other, couple
(through induced dipole-coupling) such that their resonance frequency shift compared
to well-separated particles. Thus, the presence of a tight knot is expected
to show up as a shift in resonance frequence for the particles in the knot
region. Plasmon particles can be manufactured with different resonant
scattering wavelengths. Different parts of the DNA may therefore be labelled
by particles with different resonance wavelength, allowing for detection not
only of the presence of a knot (through a shift in resonance frequency) but
also its rough location (through the absolute value of the resonance
frequency).

We point out that FRET labels could improve the knot size resolution, and
potential size changes of the knot as a function of time. Moreover,
combination of the above setup with the locally induced DNA-denaturation
as discussed in section \ref{memo}, it should be possible to observe a
decrease in the knot size when temperature or solvent conditions are changed,
due to the considerably smaller persistence length of ssDNA compared
to dsDNA.

\subsection{Target search of proteins on a DNA}

\begin{figure}
\includegraphics[width=8cm]{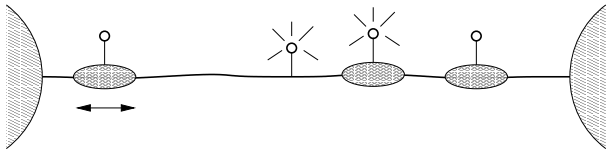}
\caption{An acceptor dye placed in the microscope focus on the DNA emits FRET
signals once a protein, that is equipped with a donor dye, comes in
close contact.}
\label{targetfind}
\end{figure}

Figure \ref{targetfind} shows a possible way to obtain information
about the target search process of DNA-binding proteins on a DNA. This is
connected to the important question about the dynamical details of how
transcription factors, that regulate gene expression, find the specific target
sequence they are supposed to bind to as efficiently as they are known to. In
the Berg-von Hippel model for target search \cite{berg}, this could be explained
by a combination of volume diffusion and one-dimensional sliding of the
proteins along the DNA while being non-specifically bound. A quantitative
study of this model has not been achieved up to date.
A large fraction of binding proteins are non-specifically bound to DNA
\cite{audun}; in that state, sliding motion is their only means of propagation.
There have even been identified
cases when each binding protein remains on the DNA during the entire target
search process \cite{mark}.

The setup from Figure \ref{targetfind} provides a possibility to obtain
quantitative single molecule information about the targeting process.
By labelling the protein with a donor dye, which has a corresponding acceptor
dye on the DNA
molecule, that is held in place by the optical tweezer, one can measure
by FRET single events when an individual protein comes within a few {\AA}
of the DNA-dye. The obtained time series can then be converted into the
desired dynamical information such as sliding diffusion constant etc, as
function of solution conditions, protein types, or DNA sequence. In
addition, the effects of DNA-knots on the target search may be studied.

Alternatively, one might dress the binding proteins by plasmon
resonant nanoparticles (compare to subsection \ref{sec:DNA_knot}).
Possible clustering effects may then be detected as shifts in the
resonance frequency spectrum of these particles. Moreover, from such
measurements there is hope to extract the local concentration
of binding proteins as a function of time, as well as interactions between
the proteins (cooperativity effects).

\subsection{Measuring protein binding using DNA translocation} 

Above discussion shows that by use of single DNA setups, DNA itself
can be probed, or used to probe its environment on the micro- and
nano-scale.  Here, we suggest one possible experiment, in which DNA
(and binding proteins) can be employed to create small, controllable
forces. The results here also allow for the measurements of
equilibrium binding properties and binding kinetics.

\begin{figure}
\includegraphics[width=8cm]{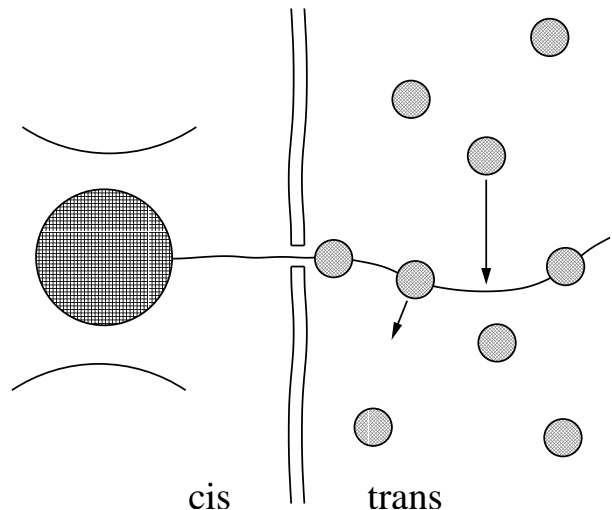}
\caption{Binding proteins that adsorb to and unbind from the part of the
biopolymer to
the right of the membrane partially rectify its translocation through 
the pore. Held in a quasistationary mode by an optical tweezer bead to
the left of the membrane, the small forces exerted by the binding
proteins can be monitored.}
\label{translo}
\end{figure}

Our example relates to binding proteins, that by reversible binding
(partially) rectify the passage of a biopolymer through a membrane nanopore
\cite{alberts}. The experimental setup we have in mind is depicted in Figure
\ref{translo}: once an end of the biopolymer is threaded through the pore,
binding proteins on the trans side
can (reversibly) bind. While bound, a protein prevents
backsliding through the pore such that the diffusive motion of the biopolymer
through the pore becomes (partially) rectified. A microbead attached to the
end of the biopolymer that is on the cis side of the pore, experiences a
net drag force towards the pore that can be measured, for instance, by
monitoring the displacement of the bead in an optical trap. The typical
force exerted onto the connected microbead in such a setup can be approximated
as a few pN, and below; compare the analysis in Reference \cite{ambme},
and the experiments reported in Reference \cite{elbaum}, for which binding
protein-mediated ratcheting was proposed as the most likely mechanism. 
The advantage of such a force transducer
may be in the possibility of a slow build-up of a small force, in
comparison to optical tweezers or similar single molecule tools that
are typically run with constant force or constant velocity protocols.
The setup connected to a sensitive force-meter such as an optical
tweezer could measure (i) the concentration of binding proteins; (ii)
the protein binding constants (iii) the size of the proteins
\cite{ambme}. An alternative to the binding proteins could be a
molecular motor such as polymerase progress along a DNA that is
threaded through the nanopore, and thus create a relatively constant
pulling speed.

When the translocating biopolymer is a flexible single-stranded DNA
the force exerted on the microbead additionally depend on: (iv)
entropic forces, which in turn depend on the persistence length; (v)
interactions between the surface and the biopolymer \cite{Ed_John}.

One may measure (vi) the unbinding rate in the following way: one moves the
bead ``backwards'' slowly until it stops (due to the presence of a
bound binding protein). When the binding proteins unbinds the bead can
again move. The average stopping time is a measure of the unbinding
rate constant.

\section{Perspective Nanobiotechnology}

During the last decade or so, single molecule methods have taken root in
disciplines like soft matter and biological physics,
and it was demonstrated in numerous experiments the basic potential and
feasibility of these techniques. By now, the time appears to be ripe to
explore new possibilities for the applications of these methods. In the present
work, we collected a number of potential experimental setups that can be
employed to both explore the physical properties of DNA and its interaction
with other molecules, as well as to utilize DNA as
sensitive probe of its environmental conditions. These setups mainly concern
the micro- and nanometre range, and may therefore be particularly useful in
small volumes such as the microdishes of gene arrays, in microreactors, or
as monitors or micromachines that are introduced in cells.

The potential applicability of these experiment designs, as far as they involve
fluorescence techniques, relies crucially on the quality of the dyes. Whereas
typical fluorophores bleach out within rather short timescales, the quantum
dots and plasmon resonant nanoparticles we mentioned earlier provide a robust
alternative, that will doubtlessly boost fluorescence techniques in small
systems.

Our proposed setups are all based on the physical properties of DNA and the
interactions dynamics with its environment; in particular, entropy and
free energy effects. In this sense, the characteristics studied here are
similar to previously suggested designed chemical molecules, whose
functionality is based on entropic units such as sliding rings
\cite{cpl,cpl1}.

Various of the proposed setups involve the fixation of the DNA molecule by
an optical tweezer. A few words on this methods are therefore in order.
Firstly, it should be kept in mind that the typical size of such beads is
of the order of a $\mu$m. For short DNA to be investigated, surface effects
due to the beads may therefore come into play that obviously decrease with
increasing chain length. Secondly, there is a tradeoff between the positional
fixing of the chain and the magnitude of the applied pulling force. Whereas
for very small forces the chain is only marginally affected and a statistical
segment close to its centre still has a large amplitude of motion, a very
large pulling force can keep the same segment almost still but changes the
statistical properties of the DNA significantly (for instance, it can become
close to force-induced denaturation). In between these two regimes, the blob
picture is a good description of the DNA chain at lower to intermediate
forces \cite{degennes}: the pulling
force $f$ then gives rise to blobs of size $\xi=k_BT/f$ in which the DNA is
undisturbed; the blobs themselves align parallel to the force vector. At
intermediate to higher forces, the worm-like chain model applies \cite{wlc}.
Depending on the effects intended to probe, different of these regimes may
be chosen.

We point out that many of the "old" single macromolecular techniques
measure quantities which are equilibrium averages over the macromolecule, for
instance by DNA pulling experiments one obtains the average behaviour
of the entire molecule. Many of the methods presented here allow for
the study of \emph{local} (nanometre to subnanometre) and \emph{dynamical}
behaviour of DNA and its local environment, thereby providing new
opportunities in the studies of subcellular behaviour and biotechnology.
In that sense, the proposed setups represent a continuation of recent
experiments such as the pulling of small RNA hairpins by optical tweezers
\cite{liphardt} and their opening and closing dynamics (molecular beacon)
as measured by fluorescence \cite{olegrev}, the polymer dynamics of
dsDNA versus ssDNA \cite{oleg}, or the persistent length and its sequence
dependence of ssDNA measured by fluorescence methods \cite{murphy,oleg1}.

We hope that the this study will inspire the design of novel DNA-based
single molecule experiments.


\bibliographystyle{unsrt}
\bibliography{refs}

\end{document}